# Imaging non-local electron transport via local excess noise

Qianchun Weng[1, 4*], Susumu Komiyama[1, 5*], Zhenghua An[2, 3†], Le Yang[2], Pingping Chen[1], Svend-Age Biehs[6], Yusuke Kajihara[4] & Wei Lu[1†]

[1] National Laboratory for Infrared Physics, Shanghai Institute of Technical Physics, Chinese Academy of Sciences, Shanghai 200083, PR China.
[2] State Key Laboratory of Surface Physics and Key Laboratory of Micro and Nano Photonics Structures (Ministry of Education), Department of Physics, Fudan University, Shanghai 200433, PR China.
[3] Collaborative Innovation Center of Advanced Microstructures, Nanjing 210093, PR China.
[4] Institute of Industrial Science, the University of Tokyo, Komaba 4-6-1, Meguro-ku, Tokyo, 153-8505, Japan.
[5] Department of Basic Science, the University of Tokyo, Komaba 3-8-1, Meguro-ku, Tokyo 153-8902.
[6] Institut für Physik, Carl von Ossietzky Universität, D-26111 Oldenburg, Germany.

*These authors contributed equally to this work.
†Email: anzhenghua@fudan.edu.cn (Z.A.); luwei@mail.sitp.ac.cn (W.L.)

**Noise is usually a hindrance to signal detection. As stressed by Landauer, however, the noise can be an invaluable signal that reveals kinetics of charge particles.[1] Understanding local non-equilibrium electron kinetics, acceleration and relaxation at nano-scale is of decisive importance for the development of miniaturized electronic devices, optical nano-devices, and heat management devices. In non-equilibrium conditions electrons cause current fluctuation (known as excess noise) that contains fingerprint-like information about the electron kinetics.[2-5] A crucial challenge is hence a local detection of excess noise and its real-space mapping. However, the challenge has not been tackled in existing noise measurements because the noise studied was the spatially integrated one. Here we report the experiment in which the excess noise at ultra-high-frequency(21.3±0.7 terahertz (THz)), generated on GaAs/AlGaAs quantum well (QW) devices with a nano-scale constriction, is locally detected and mapped for the first time. We use a sharp tungsten tip as a movable, contact-free and noninvasive probe of the local noise, and achieved nano-scale spatial resolution ($\Delta X \approx 50$ nm). Local profile of electron heating and hot-electron kinetics at nano-scales are thereby visualized for the first time, disclosing remarkable non-local nature of the transport, stemming from the velocity overshoot and the intervalley hot electron transfer. While we demonstrate the usefulness of our experimental method by applying to mesoscopic conductors, we emphasize that the method is applicable to a variety of different materials beyond the conductor, and term our instrument a 'scanning noise microscope' (SNoiM):In general non-equilibrium current fluctuations are generated in any materials including dielectrics, metals and molecular systems, whenever they are driven away from equilibrium by injection of external energies. The fluctuations, in turn, excite fluctuating electric and magnetic evanescent fields on the material surface, which can be detected and imaged by our SNoiM.**

In non-equilibrium conditions, thermal fluctuation of charge particles in any material is unavoidably enhanced to yield electromagnetic excess noise. Since nanoscopic information of the non-equilibrium kinetics of charge particles is imprinted in the excess noise, it is a unique physical entity, through which non-equilibrium phenomena inside the material can be nanoscopically studied. This unique property of excess noise, however, has not been exploited for the study of materials up to the present time. In this work we carry out nanoscopic mapping of the local excess noise on mesoscopic semiconductor devices, and demonstrate that the method provides us with a novel and unique metrology of matters.

Current-induced excess noise has been extensively studied both in diffusive[6-8] and coherent[9,10] conductors. However, its nanoscopic profile has not been studied because ohmic electrodes[6], microwave receivers[7,8] and on-chip quantum devices[9,10] used for probing the noise have been neither local nor movable. Very recently, local detection has been done for excess noise[11] and for thermal noise[12], respectively, by making use of a semiconductor nano-wire and an atomic-scale nitrogen vacancy as noise sensors, but the sensors were not movable. Another potential restriction to the space resolving power has been that the highest noise frequency was limited to a few hundreds of gigahertz (GHz).[8-10] Theoretically, Gramespacher and Büttiker suggested to make use of a scanning tunneling microscope (STM) tip as a local and movable noise probe,[13] and showed that the detected noise intensity directly relates to an effective local non-equilibrium distribution function of the charge carriers below the tip. Our work is the experimental realization of the theoretical suggestion in the limit of weak tip-conductor coupling (with vanishing tunneling current) at an ultra-high frequency range (21.3±0.7 THz): Excess noise is non-invasively probed and mapped with a contact-free sharp tungsten tip coupled capacitively/inductively to the conductor (Fig.1, see Methods section 'Scanning Noise Microscope (SNoiM)').

The experimental set-up resembles passive scattering-type near-field optical microscopes (s-SNOM) developed for the study of thermally generated fluctuating electromagnetic evanescent fields on the material surface.[14-19] Our SNoiM is different, however, from many of those instruments[14,17,18] in that excess noise is detected by injecting electric energy into the targeted small nano-device by transmitting current while both the sample substrate and the metal tip are kept at room temperature. The noise detection in this condition is made possible by making use of an ultra-high sensitive cryogenic sensor called CSIP.[20, 21]

We study small conductors of quasi two-dimensional electron gas (2DEG) with a submicron-wide constriction (Fig. 1b), fabricated in a GaAs/AlGaAs quantum structure with the 2 DEG layer buried 13 nm below the surface (see Methods section 'GaAs/AlGaAs quantum well structure and devices'). Figure 1c shows a two-dimensional image of the excess noise intensity obtained by modulating the tip height at a bias voltage of $V_b$ = +6V. A distinct region of intense noise shows up around the constricted region, and the region largely expands towards downstream side of the constriction (the positive voltage side). Figure 1d shows that the feature of the asymmetric expansion is reversed in the opposite polarity of bias voltage ($V_b$ = -6V), assuring that this is an intrinsic effect, not being an artifact due to sample inhomogeneity. We have studied fundamental characteristics of the noise in additional experiments, and confirmed that the noise signal detected here is due to the non-equilibrium electrons agitated by the current, and not due to lattice heating (see Methods section 'Fundamental characteristics of the detected noise'). Figure 1e shows that the amplitude of excess noise rapidly decays as the tip is moved away from the surface. This shows that the detected noise fields are the evanescent waves decaying rapidly with increasing the distance from the conductor surface, which is consistent with the theoretically predicted characteristics of the electromagnetic density of states (EM-LDOS),[22,23] which connects the fluctuating evanescent fields to the current fluctuation inside the material (see Methods section 'Electromagnetic local density of states (EM-LDOS)').

A first simple interpretation of the remarkable non-local nature of the transport (Figs. 1b,c) is obtained by considering the kinetics of electrons around the constricted region. When electrons approach and enter the constricted region (from the negative-voltage side), they are accelerated by the electric fields. When the accelerated electrons exit from the constriction (to the positive-voltage side), they, in turn, cool down by releasing their excess energies to the lattice. It is important that the cooling-down process is relatively slow because the energy relaxation time due to electron-phonon interaction is relatively long ($\tau_{e-ph} \approx$ 1.1ps). (For quantities relevant to the transport, see Methods section 'Important quantities relevant to the transport').The energetic electrons, on the other hand, drift down the constricted region at very high velocities, $v_d \approx$ (1.9~2.1) x$10^5$ m/s (velocity overshoot).[8,24,25] It follows that the energetic (or hot) electrons drift a large distance, $L_{drift,e-ph} = v_d \tau_{e-ph} =$ 210~230 nm, without being equilibrated with the lattice.

For deeper understanding we note that the inequality relation[7]

$$L_{e-e} < \Delta X \ll L_{e-ph} \qquad (1)$$

holds in our conductor. Here, $\Delta X \approx$ 50 nm is the size of 2DEG area probed by the tip, which gives the effective conductor size in SNoiM, and $L_{e-e} \approx$ 39 nm and $L_{e-ph} \approx$ 150 nm are the diffusion lengths of electrons, respectively, for the intra-electron-system equilibration and for the inter-electron-phonon equilibration. (See Methods section 'Shot noise and hot electrons'.) It follows that the non-equilibrium electrons are in quasi-equilibrium within the electron system, but are out of equilibrium to the lattice system. Hence, the non-equilibrium distribution is approximately characterized by an effective electron temperature $T_e$ (higher than the lattice temperature $T_L$).[3,5,6] It follows that the excess noise can be analyzed by using the theory of thermal noise (Johnson-Nyquist noise),[26,27] and $T_e$ can be estimated from the experimentally detected noise intensity as indicated on the scales of Figs. 1c,d (see Methods section 'Analysis of the shot noise and estimation of $T_e$').The hot-electron shot noise is associated with ultrafast kinetic processes of dissipation such as energy relaxation to the lattice, intervalley scattering, inelastic electron-electron scattering and inelastic impurity scattering.

Evolution of the hot electron distribution with increasing $V_b$ from 0.5V up to 8V is displayed in Figs. 2a-f. Hot electron distribution is recognized when $V_b$ exceeds ~0.8V, and it grows with further increasing $V_b$ (Figs. 2a-c). The asymmetric expansion of the hot electron distribution towards the downstream side of the constriction becomes distinct for $V_b$ above ~ 4.0V (Fig. 2d,e). An even more striking feature shows up at higher values of $V_b$ (Fig.2f); namely, for $V_b$ > 6V the noise intensity increases as the electrons move away from the constricted region, creating a hot spot at a distance 200~300 nm away from the constriction. In Fig. 2g, this is demonstrated by the development of a distinct maximum peak of noise intensity at the shifted position (y ≈ 250 nm) with increasing $V_b$.

Additional experiments show that the region of high electric fields is confined in a narrow area around the constricted region and that the expanded hot electron distribution (Figs. 2d-f) distinctly exceeds the high-electric-field region (see Method section 'Electrostatic-potential distribution studied by the scanning gate microscopy (SGM)'). Figure 2h shows that the current $I$ passing through the device is saturated with increasing $V_b$ above ~4.0 V. The noise intensity at the constriction is saturated in the $V_b$-range where $I$ is saturated, but the one at the hot spot continues to increases without exhibiting saturation.

The electric field at the center of the constricted region (y=0), $E_c$, is estimated to reach ≈120 kV/cm at $V_b$ =9.0V(see Methods section 'Estimation of the electric field'). It is well known that, hot electrons transfer to the upper satellite valleys at high electric fields in GaAs (see Methods section 'Valley transfer').[8,28,29] The effective mass of electrons in the upper valleys is substantially larger than in $\Gamma$-valley, so that the conductance is thereby significantly reduced, giving rise to several remarkable effects including Gunn effect.[28] In this work, the valley transfer enhances remarkably the non-local nature of the electrical conduction, like the expansion of the hot electron distribution to the downstream side of the constriction or the occurrence of hot spot outside the constriction, as described in the followings.

According to existing experiments[8] and the Monte Carlo simulation,[29] the $\Gamma$-to-X valley transfer (the splitting energy $\Delta\varepsilon_{\Gamma X}$ ≈ 550 meV) sets in at $E_c$ ~ 15 kV/cm ($V_b$~4.0 V). Hence the saturation of current I (Fig. 2h) is interpreted as being caused by the valley transfer. We note that in the experiments the remarkable expansion of the hot electron distribution to the downstream side of the constriction along with the occurrence of hot spot outside the constriction take place when $V_b$ exceeds 4.0 V; viz., the remarkable non-local features are a consequence of the valley transfer.

Our interpretation is schematically represented in Figs. 3. When electrons approach the constriction, electrons are heated up while they are increasingly transferred to the X-valleys (Fig. 3b, left): Hence, the average energy, $<\varepsilon_\Gamma>$, increases while the relative population, $n_\Gamma$, of the electrons in the $\Gamma$-valley decreases (Fig. 3c). Right at a center of constriction, y ≈ 0 (Fig. 3c, center), $<\varepsilon_\Gamma>$ reaches maximum while $n_\Gamma$ falls to minimum, where the minimum value, $n_{\Gamma,c}$, is estimated to be 0.3~0.23 for $V_b$ = 6.0~9.0 V ($E_c$= 60~120 kV/cm). At the constriction, the current is substantially carried by the electrons in the $\Gamma$-valley although $n_{\Gamma,c}$, is low because the drift velocity, estimated to be $v_{d,\Gamma}$ ≈ (1.9 ~2.1) x $10^5$ m/s, is much higher than the velocity of X-valley electrons. The value of $v_{d,\Gamma}$ largely exceed the steady-state saturated values in long channels, (8.0~9.0)x$10^4$m/s. This is a general feature known as the velocity overshoot, expected in short channel devices. On the downstream side of the constriction(Fig. 3d, right), $<\varepsilon_\Gamma>$ decreases while $n_\Gamma$ increases because the electric field E decreases with increasing y and the electrons in the X-valleys are scattered back to the $\Gamma$-valley. The value of $n_\Gamma$ rapidly recovers the low-field value nearly tracing the profile of electric field strength because of the high inter-valley scattering rate. In contrast, $<\varepsilon_\Gamma>$ decreases much more slowly with increasing y (Fig. 3c) due to threefold mechanisms. First, the electron-phonon energy relaxation is a slow process as already discussed. Secondly, the drift velocity of electrons in the Γ-valley is high due to velocity overshoot. Thirdly, during the cooling down process, back transfer of electrons from the X-valleys supplies very high energy electrons (~500 meV) in the Γ-valley. The significant expansion of the hot electron distribution towards the downstream is hence the consequence of the velocity overshoot and the valley transfer of hot electrons.

The occurrence of hot spot at a location outside the constricted region is interpreted as follows. The excess noise is generated substantially by the Γ-valley electrons because its average kinetic energy $<\varepsilon_\Gamma>$ is much higher than that of X-valley electrons $<\varepsilon_X>$.[29] The noise power density is an increasing function of both $<\varepsilon_\Gamma>$ and $n_\Gamma$, so that $n_\Gamma<\varepsilon_\Gamma>$ may give a rough measure of the noise intensity (see Methods section 'Valley transfer'). As illustrated in Fig. 3c, $n_\Gamma<\varepsilon_\Gamma>$ forms a maximum peak in a region away from the constriction, which reproduces the experimental feature (Fig. 2g). We also suggest that the saturation of noise intensity at the middle of the constriction (Fig. 2h, y=0) is a consequence of the cancellation of the increase in $<\varepsilon_{\Gamma,c}>$ and the decrease in $n_{\Gamma,c}$ with increasing $V_b$ ($E_c$). The saturation-less increase of the noise intensity at the hot spot (Fig. 2h, y=250 nm) is readily understood by noting that $n_{\Gamma,c}$ quickly recovers a value $n_\Gamma = 0.7$~$0.8$ at the hot spot and $<\varepsilon_\Gamma>$ increases with increasing $V_b$ ($E_c$).

The effective electron temperature is estimated to reach $T_e$ = 2,200~2,500K ($<\varepsilon> \equiv (3/2)k_B T_e$ = 280~320 meV) at the hot spot (Figs. 2f-h and Fig. 3a). We note that the estimated $T_e$ is an average value of the effective electron temperature of the Γ-valley, $T_{e,\Gamma}$, and that of the X-valleys, $T_{e,X}$, where $T_{e,\Gamma} \gg T_{e,X}$ is expected. A rough estimate suggests $T_{e,\Gamma}$= 2,800~3,200K ( $<\varepsilon_\Gamma>$= 350~400 meV), which are comparable to the values, $<\varepsilon_\Gamma>$= 300~570 meV, predicted by the Monte Carlo simulation.[29]

In summary, we visualized hot electrons in the nano-scale for the first time at the platform of GaAs/AlGaAs devices by mapping the real-space excess noise in the presence of current. Our work opens up the possibility of a new metrology of matters, in which non-equilibrium phenomena on the nanometer scales can be probed and imaged in terms of the local current fluctuation (excess noise).

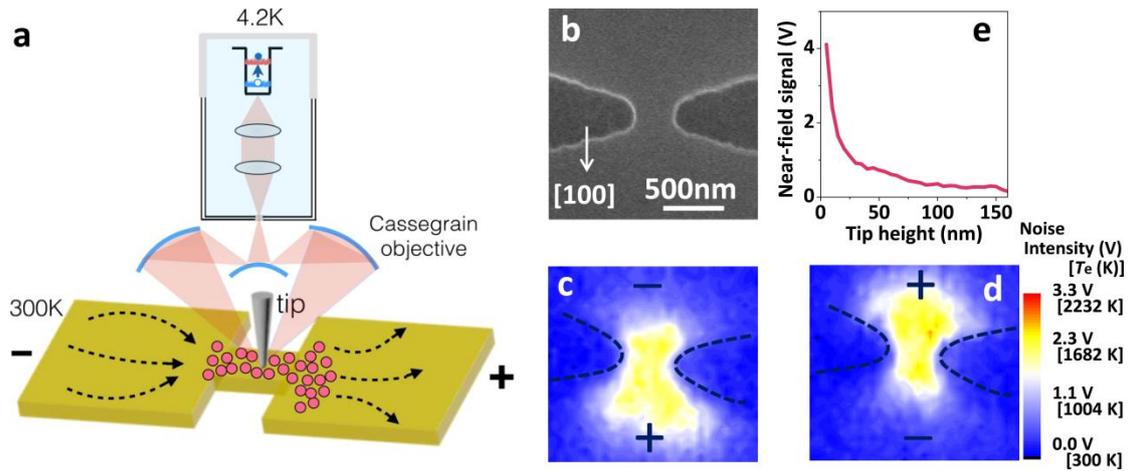

**Figure 1|** Nano-scale mapping of ultra-high frequency excess noise (21.3±0.7 THz) with scanning noise microscope (SNoiM). a, Schematic representation of the experimental setup of SNoiM. b, SEM image of the nano-device with a constriction fabricated in a GaAs/AlGaAs quantum well (QW) structure. c, d, Two-dimensional real-space images of the excess noise intensity, obtained with the opposite bias polarities (±6V). The images are obtained by modulating the tip height (See Methods). The color scale is given by the signal amplitude V (output of the lock-in amplifier) as well as by the effective electron temperature $T_e$ (K) (see Methods section 'Estimation of $T_e$'). The electric field at the constriction is estimated to be $E_c \approx 62$ kV/cm for $V_b$=6.0V. e, The decay profile of the noise signal with increasing the tip-height, taken at the center of constriction with $V_b$= 6V. The signal is taken by modulating $V_b$ between 0 and 6V with the tip height fixed at 10nm (See Methods).

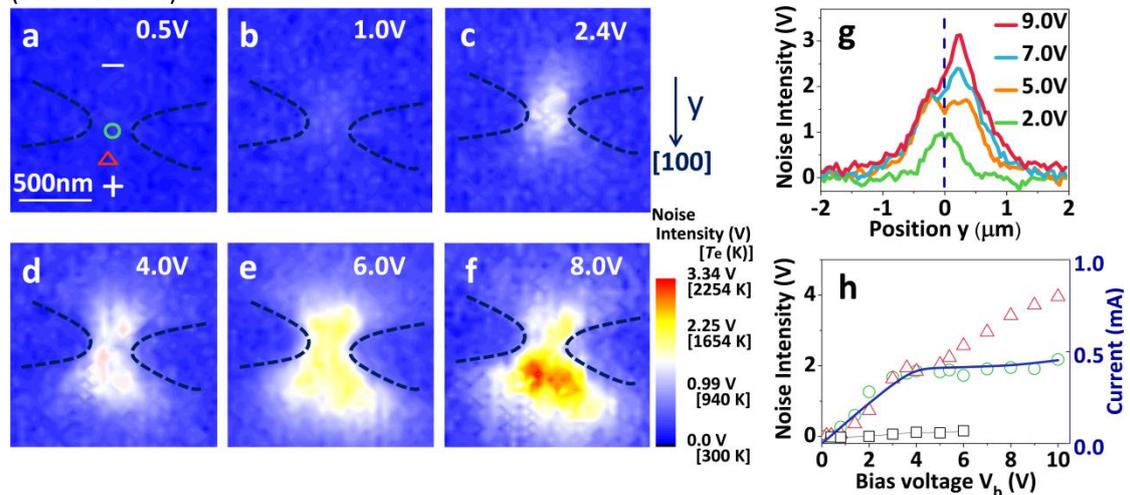

**Figure 2|** Evolution of the excess noise in real-space with increasing the bias voltage, $V_b$. a through f, 2D images. **g**, One dimensional profile of the excess noise intensity in the y-direction (the right side of Fig. 2c), where y = 0 corresponds to the center point of the constriction marked by the green circle ○ in Fig.1 a. The crystallographic orientation is y//[100]. **h**, The solid line shows the current vs. $V_b$ curve. The data points of ○ and △ show, respectively, the excess noise intensity at the positions marked by ○ (y=0) and △ (y=+250 nm) in Fig. 2a against $V_b$. The black squares □ plot the signals obtained by modulating bias voltage without the tip, showing that the far-field component is absent.

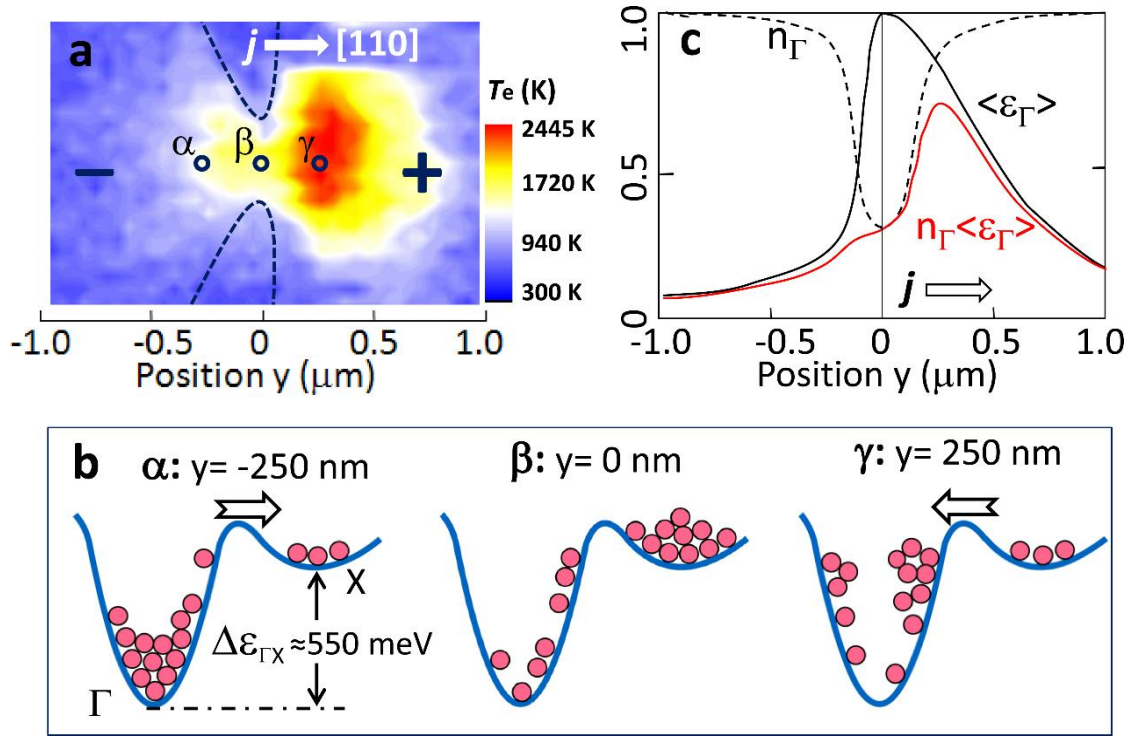

**Figure 3|** Hot electron kinetics in the vicinity of constricted region. a, Experimentally obtained image of the hot electron distribution for $V_b$=9.0V ($E_c$≈120 kV/cm), taken in a constriction device different from the one shown in Figs. 1 and 2. b, Schematic representation of the relative population of electrons in the Γ- and the X-valleys, respectively at positions α, β, and γ in Fig. 3a. c, A sketch of our interpretation on the variation of $n_\Gamma$<$\varepsilon_\Gamma$> and $n_\Gamma$<$\varepsilon_\Gamma$> on the y-axis for $V_b$= 6.0~9.0 V, where $n_\Gamma \equiv N_\Gamma/N$ ($N_\Gamma$; the density of Γ-valley electrons, N; the total density of electrons) is the relative population of Γ-valley electrons, <$\varepsilon_\Gamma$> is the average kinetic energy of Γ-valley electrons, and $n_\Gamma$<$\varepsilon_\Gamma$> is a measure of the excess noise intensity.

**Acknowledgements** We thank the research support from the National Natural Science Foundation of China under grant Nos. 11427807/11634012/11674070 and from Shanghai Science and Technology Committee under grant No. 16JC1400400. S.K. acknowledges support by the Chinese Academy of Sciences Visiting Professorships for Senior International Scientists. Q. W. is now an International Research Fellow of the Japan Society for the Promotion of Science. Y. K. acknowledges support from Collaborative Research Based on Industrial Demand by Japan Science and Technology Agency. Z.A. thanks Profs. L. Zhou and S.W.Wu for helpful discussions. Part of the experimental work was carried out in Fudan Nanofabrication Laboratory.

**Author contributions** S.K., Q.W. and Z.A. conceived the idea, analyzed the data and co-wrote the manuscript with constant discussion with W.L. The SNoiM was constructed by Q.W. and Z.A. following the advice and the design of Y.K. and S.K. The experiments were carried out by Q.W. on nano-devices fabricated by L.Y. in the wafers grown by P.C. The simulation calculation of EM-LDOS was done by S.A.B. The research projects were co-supervised by Z.A. and W.L.

# Methods

Scanning Noise Microscope (SNoiM).
GaAs/AlGaAs quantum well structure and devices.
Fundamental characteristics of the detected noise.
Electromagnetic local density of states (EM-LDOS).
Important quantities relevant to the transport.
Shot noise and hot electrons.
Estimation of $T_e$.
Electrostatic-potential distribution studied by the scanning gate microscopy (SGM).
Estimation of the electric field.
Valley transfer.

# Extended Data Figures

## Methods

**Scanning Noise Microscope (SNoiM).** The noise frequency studied in this work is so high that standard electronic circuits cannot be applied. A technique of near-field optics is therefore applied (Fig.1a), where current fluctuations in the conductor induce fluctuating currents in the tungsten tip through the capacitive/inductive coupling, which, in turn, causes the tip to emit radiation in the far field, and the emitted radiation is detected. In other terms, fluctuating currents in the conductor generate fluctuating electric and magnetic evanescent fields in the proximity of the conductor surface (see Methods section 'Electromagnetic local density of states (EM-LDOS)'): The evanescent fields are scattered by the tungsten tip, collected with a Cassegrain objective (numerical aperture, NA=0.61) and led to an ultra-highly sensitive two-level detector CSIP operated at 4.2 K.[20,21] The CSIP is a home-made detector with a spectral band of 14.1±0.5 μm(21.3±0.7 THz) and which is by a factor of few orders of magnitude more sensitive than a commercially available mercury cadmium telluride (MCT) detector.

In order to extract the near-field evanescent-field component out of unwanted background radiation, the tip height $h$ is modulated with an amplitude 100 nm at a frequency Ω= 5 Hz and the detector signal is demodulated at the fundamental frequency (5 Hz). When necessary, measurements are made also by modulating the bias voltage (with square waves alternating between 0 and $V_b$) while keeping the tip height fixed at the bottom position ($h \approx 10$ nm).

In many of the passive scattering-type near-field optical microscopes (s-SNOMs), commercially available MCT detectors are applied, and either the sample[14] or the tip[17,18] is heated up so that the thermal fluctuation in the sample is intensified to a detectable level. Our microscope system is different from those s-SNOMs but similar to more sensitive s-SNOMs reported in Refs. 15,16 and 19, with which thermal evanescent fields on metals and dielectrics can be detected without heating up the sample or the tip owing to the ultra-high sensitivity of the CSIP detector. In this work, the sample and the probe are mounted on room-temperature stages and placed in the room-temperature atmosphere, so that the temperatures of the sample and the probe tip are equal to the room temperature; viz., $T_{sample}=T_{tip}= T_{Room} \approx$ 300K. Another important factor is that the effective temperature $T_{Rad}$ of the background radiation impinging on the sample is lower than $T_{Room}$ whereas the sample is placed in the room-temperature atmosphere:[16] $T_{Rad} \approx$230K is estimated in the SNoiM used. The inequality condition $T_{sample}> T_{Rad}$ is a fundamental prerequisite for the passive measurements either of thermally-activated or current-induced excess noise.

The spatial resolution ΔX is primarily determined by the radius of curvature of the tip apex, as reported for a s-SNOM system similar to our SNoiM.[30] In the SNoiM used in this work, the radius of curvature is ~50 nm and ΔX≈ 50 nm is obtained as discussed in Methods section 'Fundamental characteristics of the detected noise'.

**GaAs/AlGaAs quantum well structure and devices.** Extended data Fig.1 shows the heterostructure used in this work. The crystal with (100) plane is grown by molecular beam epitaxy (MBE). Quasi two-dimensional electron gas (2DEG) layer is provided by delta-doping of Si in the 35 nm-wide QW 13 nm below the surface. The electron density and the electron mobility at room temperature are derived from the Hall-effect measurement to be $n_{2D}$= 1.10 x $10^{16}$m$^{-2}$ (n= 3.3 x $10^{24}$m$^{-3}$) and µ=0.167 m$^2$/Vs. Extended data Fig.2 shows an example of GaAs nano-constriction devices fabricated in the GaAs/AlGaAs QW structure via standard electron beam lithography (EBL). A 2DEG channel is patterned by wet-mesa etching (depth~ 80nm): A narrow constricted 2DEG region is connected to the drain and the source contacts through 2DEG leads of width $W_{lead}$= 20µm and length $L_{lead}$= 190 µm. Ohmic contacts (source and drain) are prepared by alloying with AuGe. A cross of 20µm-width gold layers (120 nm thickness) is deposited to serve as the position marker for identifying the focal spot of the SNoiM. The gold stripes of the cross are electrically floating and isolated from the 2DEG. The right-hand side of Extended data Fig. 2 shows the central region including a constriction. The current flows along [100] or [110] direction. Several samples with differently shaped constrictions are fabricated (Extended data Fig. 6).

**Fundamental characteristics of the detected noise.** As displayed in Extended data Figs.3, the bias-polarity dependent asymmetric profile of excess noise distribution, such as shown in Figs. 1c, d, Figs.2 e, f, and Fig. 3a, are found in other devices with modified shape of constricted regions and in a device along different crystallographic orientation [110] in the high-bias condition. While all these images of excess noise are taken by modulating the tip height, we have confirmed that similar images are obtained also when bias voltage $V_b$ is modulated while the tip height is fixed at a position close to the surface. Hence we can be convinced that the noise signal occurs only when the tip is placed close to the surface in the presence of current: This feature provides compelling evidence to show that the noise studied here is due to the current-induced fluctuating electric and magnetic evanescent fields.

When the tip is removed, the SNoiM serves as a thermometry microscope for the far-field radiation of a wavelength ≈14.1±0.5 µm. Extended data Figs.5 compares thermometry images taken with and without bias voltage. No discernible difference is recognized. Hence, a possible temperature rise is inappreciably small (< 2.0K) everywhere in the substrate including the region of the nano-constriction device and the 2DEG leads. The very small lattice heating is readily understood by noting that the lattice heat capacity[31] is larger than the electron heat capacity (≈ 3n$k_B$) by a factor as large as six orders of magnitude; hence, without invoking detailed discussion we can understand that the rise in the lattice temperature is far smaller than the rise in the effective electron temperature(see Methods section 'Shot noise and hot electrons'). By assuming that the Joule heating, P= $V_b$ x I, occurs in the constricted 2DEG region and that the heat diffusively propagates isotropically in the crystal, we can roughly estimate the lattice temperature on the hemisphere of radius r (> 300 nm) with the constriction at the center as T(r) = $T_0$ + ΔT(r) with ΔT(r) = {P/(2π$K_L$)}/r, where $T_0$ is room temperature and $K_L$ ≈ 50 W/m·K[31] is the lattice thermal conductivity of GaAs. We have ΔT ≈ 1.0K for r = 15 µm from P = $V_b$ x I ≈ 4 mW, which is consistent with the null result of Extended data Fig. 4(D). Non-equilibrium phonon distribution (hot phonon distribution)[32] is possible to occur in general, but the effect may be far less remarkable than the effect due to hot electrons in this work.

Extended data Figs.5 shows that the spatial resolution determined by the baoundary transition width is ΔX ≈ 50 nm.

**Electromagnetic local density of states (EM-LDOS).** In SNoiM, fluctuating currents in the electron system generate fluctuating electric and magnetic fields in the vicinity of conductor surface, which are scattered by the tip. The electron system under study is out of thermal equilibrium. Nevertheless, It is reasonable to start discussion by considering a material in thermal equilibrium at temperature $T$. On its surface, electric and magnetic fields are generated by thermally activated current fluctuations. If the material is given by a half space with a flat interface, the local energy density of the thermally induced electromagnetic fields with angular frequency $\omega$ at a position of distance $h$ from the material surface is expressed as[22,23]

$$u(h,\omega,T) = \rho(h,\omega)[\hbar\omega/\{\exp(\hbar\omega/k_B T)-1\}] \qquad M(1)$$

with the Planck constant $\hbar$ and the Boltzmann constant $k_B$. Here $\rho(h,\omega)$ is the material-specific electromagnetic local density of states (EM-LDOS), theoretically derived by using the fluctuation-dissipation theorem, and is expressed as a function of the Fresnel reflection coefficients. It is to be mentioned that M(1) is reduced to the energy density of black-body radiation (given by Planck's radiation theory) when $\rho$ is replaced with the vacuum EM-LDOS $\rho_{vac}(\omega)=\omega^2/(\pi^2 c^3)$.

Extended data Fig.6 shows the EM-LDOS of the GaAs/AlGaAs quantum well (QW) structure shown in Extended data Fig.1, $\rho_{n-GaAs}$, along with the reference data of gold, $\rho_{Au}$, as a function of wavelength $\lambda = 2\pi c/\omega$ ($c$; the light velocity). The EM-LDOS of the GaAs/AlGaAs QW arises primarily from the 2DEG in the QW, since the surface phonon polariton resonances of the GaAs and the AlAs sub-lattices are away from the target wavelength, $\lambda_{target} \approx 14.1$ μm. Additional calculation shows that at $\lambda = \lambda_{target}$ the phonon contribution is less than 10% of that of electrons. The plasma wavelength, $\lambda_{plasma}= 17.5$ μm, is longer than $\lambda_{target}$, and the absolute amplitude of EM-LDOS $\lambda = \lambda_{target}$ is smaller than that of gold by an order of magnitude. It follows that thermally activated electromagnetic evanescent fields due to the electron system in the QW are not detectable in the experiment whereas those in gold have been detected in a system similar to our SNoiM.[15,16,19] Extended data Fig.7 shows the data of thermal evanescent fields obtained on gold at T=300K with our SNoiM.

When the 2DEG is out of thermal equilibrium, the local energy density of electromagnetic fields can be approximately evaluated by replacing $T$ with $T_e$ in relation M(1), because the non-equilibrium electron distribution in this work is approximated by the hot electron distribution described by an effective electron temperature $T_e$ (see Methods sections 'Shot noise and hot electrons' and 'Estimation of $T_e$'). Although the lattice temperature is not elevated, this procedure may be valid because the contribution from the electron system dominates.

According to the theory (not shown), $\rho_{n-GaAs}(h,\omega)$ at $\lambda_{target}$ decreases with increasing h from 10 nm to 50 nm by a factor of ten, which is consistent with the experimentally found rapid decay of the noise signal with increasing $h$ (Fig. 1e).
For later analysis we also mention that the contribution of 2DEG to the EM-LDOS is theoretically shown to monotonicaly increase with increasing the electron density n in a range of $(0.7\sim3.0)\times10^{24}/m^3$: The dependence is roughly approximated by $\rho_{n-GaAs,e} \propto n^{1.5}$.

**Important quantities relevant to the transport.** The Fermi energy, the Fermi velocity and the thermal velocity of electrons in the QW (Γ-valley) are evaluated, respectively, to be $E_F$= 119 meV, $v_F$= $(2E_F/m_\Gamma^*)^{1/2}$ = 7.3 x $10^5$ m/s and $v_T$= $(3k_BT/m_\Gamma^*)^{1/2}$ = 4.2 x $10^5$ m/s (T=300K), where $m_\Gamma^*$= 0.078 $m_0$ ($m_0$: the free electron mass) is the effective mass of electrons in the Γ valley with the density n= 3.3 x $10^{24}m^{-3}$.[33] The mean free path of electrons, $l$ = $v_F\tau$ = 55 nm with the momentum relaxation time $\tau$= $m_\Gamma^*\mu/e$ = 75 fs ($\mu$=0.167 $m^2$/Vs), is larger than the QW width, W = 35 nm, but the dephasing length $l_\varphi$ is smaller than W. It follows that the electron system in the QW is regarded as semi-classical quasi two-dimensional electron gas (2DEG).

The energy relaxation time due to electron-electron scattering, $\tau_{e-e}$, is expected to be $\tau_{e-e}\approx\tau$ = 75 fs because the density of electrons is nearly equal to that of the ionized donors. The relaxation time, $\tau_{e-e}\approx$ 75 fs, is supposed not to vary significantly when electrons are heated (within the Γ valley) because the increase of the final density of states for scattering tends to compensate the reduction of the Coulomb scattering probability.

Energy dissipation of heated electrons to the lattice is dominated by the emission of longitudinal optical (LO) phonons. In the process of thermalization with the lattice, an average hot electron emits several LO-phonons because the mean energy of hot electrons in this work is <ε> = 200 ~ 300 meV while the LO-phonon energy is $\hbar\omega_{LO}$ = 36 meV so that {(e-1)/e}<ε> = (3.5 ~ 5.3) x $\hbar\omega_{LO}$ (e; the base of natural logarithm). The rate of LO-phonon emission is known to be $1/\tau_{LO}\approx$ 1/130 fs for GaAs in the limit of low electron density,[34] but in the heavily doped QW studied in this work we roughly estimate $1/\tau_{LO,HD}\approx$ 1/260 fs considering the screening effect.[35] Hence the energy relaxation time due to electron-phonon interaction is estimated to be $\tau_{e-ph}\approx$ (3.5 ~ 5.3) x $\tau_{LO,HD}\approx$ 0.9~1.4 ps. For simplicity we apply $\tau_{e-ph}\approx$ 1.1 ps as a representative value for the discussion in the text.

The drift velocity <$v_d$> of electrons passing through the constricted region is directly derived from the magnitude of current I by using the relation <$v_d$>= I/($n_{2D}eW_e$) with $n_{2D}$= 1.10 x $10^{16}m^{-2}$, where e is the unit charge and $W_e\approx$ 250 nm is the effective width constriction width. For instance, <$v_d$>$\approx$ 1.1 x $10^5$ m/s is derived from I ≈ 0.43 mA in the saturated current regime for $V_b$> 4.0 V (Fig. 2h). In the high $V_b$ range (> 4.0V), electrons are populated not only in the Γ-valley but also in the X-valleys. See Methods section 'Valley transfer' for the discussion below. The Γ-valley electrons drift much faster than the X-valley electrons because of the lighter effective mass, and are primarily responsible for the generation of excess noise. Noting the relative population (Extended data Fig. 9) and different effective masses, the drift velocity of electrons in the Γ-valley is estimated to be $v_{d,\Gamma}\approx$ (1.9~2.1) x$10^5$ m/s.

**Shot noise and hot electrons.** Letting D be the diffusion constant, the diffusion length of electrons for the intra-electron-system thermalization is given by $L_{e-e}$ = $(D\tau_{e-e})^{1/2}$, and the diffusion length of electrons for the inter-electron-phonon thermalization by $L_{e-ph}$ = $(D\tau_{e-ph})^{1/2}$. Here D =<v>$^2\tau_m$/2 with <v> the average speed of electrons and $\tau_m$ the momentum relaxation time. In general, D is a function of the electric field E.[36] In the GaAs/AlGaAs QW structure studied in this work, the electron density is relatively high, so that we can assume D to be substantially unchanged against $V_b$ because the increase in $<v>^2$ with increasing E is relatively slow, which is nearly compensated by the decrease in $\tau_m$. Hence we estimate $D \approx v_F l/2 = 2.01 \times 10^{-2}$ m$^2$/s by taking $<v> = v_F$ and $\tau_m = \tau$, from which $L_{e-e} = (D\tau_{e-e})^{1/2} \approx 39$ nm, $L_{e-ph} = (D\tau_{e-ph})^{1/2} \approx 150$ nm are obtained, which yield inequality relation (1)

$$L_{e-e} < \Delta X \ll L_{e-ph}.$$

The physical implication of Relation (1) is elucidated in detail below. Current-induced excess noise in conductors is called shot noise. It is known that dephasing of electrons does not affect shot noise, but energy relaxation with the lattice suppresses shot noise: The shot noise vanishes if the electron system is thermalized with the lattice system in large conductors: Hence the last half of the inequality relation, $\Delta X \ll L_{e-ph}$, assures the presence of shot noise in our conductor. In general, no universal system-independent expression is available for the excess noise if the electron system is far away from equilibrium because the characteristics of the noise depend on the given particular nonequilibrium conditions. In our work, however, simple analysis is possible based the "hot electron picture" supported by the first half of the inequality relation, $L_{e-e} < \Delta X$.

In conventional measurements of thermal noise of a conductor (resistance R), fluctuations are given by the well know formula $P_I = 4k_B T/R$ in the limit of low frequency $\hbar\omega \ll k_B T$. For the hot-electron shot noise, we would have

$$P_I = 4k_B T_e/R. \qquad M(2)$$

by replacing $T$ with $T_e$. In our work, however, M(2) is not applicable because the noise photon energy ($\hbar\omega \approx 82$ meV for $\omega \approx 2\pi \times 21.3$ THz) is not vanishingly small compared to the thermal energy $k_B T_e = 26 \sim 200$ meV. We have to take the general formula given by Nyquist,[27] which, after replacement of $T \to T_e$, gives

$$P_I = 4[\hbar\omega/\{\exp(\hbar\omega/k_B T_e) - 1\}]/R. \qquad M(3)$$

We note in addition that the noise is probed through the detection of fluctuating electric and magnetic evanescent fields generated on the conductor. So, the detected noise signal, $V_{SNoiM}$, is proportional not only to $P_I$ but also to the EM-LDOS, $\rho(h,\omega)$, of the GaAs/AlGaAs QW structure (see Methods section 'Electromagnetic local density of states (EM-LDOS)'); hence, the noise signal is expressed as

$$V_{SNoiM} \propto u(h,\omega,T_e) = \rho(h,\omega)[\hbar\omega/\{\exp(\hbar\omega/k_B T_e) - 1\}], \qquad M(4)$$

which reproduces M(1) for $T = T_e$, representing the local energy density of fluctuating electric and magnetic fields at distance h from the surface and at temperature $T_e$.

**Estimation of $T_e$.** Estimation can be made by comparing the noise intensity, $V_{SNoiM}$, with that of the thermal noise intensity obtained in the absence of current, $V_{SNoiM,I=0} \propto u(h,\omega,T) = \rho(h,\omega)[\hbar\omega/\{\exp(\hbar\omega/k_B T) - 1\}]$ ($T = 300$K): To be explicit, $T_e$ is derived from the ratio, $V_{SNoiM}/V_{SNoiM,I=0} = \{\exp(\hbar\omega/k_BT)-1\}/\{\exp(\hbar\omega/k_BT_e)-1\}$. Unfortunately, however, the noise signal is indiscernibly small in a low $V_b$-range where $T_e$ is not substantially higher than $T$ (Figs 2a) because the EM-LDOS of the QW structure, $\rho_{n\text{-}GaAs}$, is small (Extended data Fig.6). Fortunately, however, the EM-LDOS on gold, $\rho_{Au}$, is large enough to yield detectable signal $V_{SNoiM,I=0}^{Au}$ as shown in Extended Data Fig.7. We can estimate $T_e$, without using any adjustable parameters, from the ratio $V_{SNoiM}/V_{SNoiM,I=0}^{Au}= (\rho_{n\text{-}GaAs}/\rho_{Au})\times\{\exp(\hbar\omega/k_BT)-1\}/\{\exp(\hbar\omega/k_BT_e)-1\}$, where $\rho_{n\text{-}GaAs}/\rho_{Au} \approx 0.10$ is theoretically obtained at $\lambda = \lambda_{target}$ (Extended data Fig. 6).

It is to be mentioned that $T_e$ differs from the familiar "noise temperature $T_n$" widely discussed in the literatures:[8] $T_n$ is defined by the temperature at which the black-body yields equivalent noise; In this work, $T_n$ and $T_e$ are related to each other through the equation $(\rho_{n\text{-}GaAs}/\rho_{vac})[\{\exp(\hbar\omega/k_BT_n)-1\}]/\{\exp(\hbar\omega/k_BT_e)-1\}] = 1$ with $\rho_{vac}=\omega^2/(\pi^2 c^3)$.

**Electrostatic-potential distribution studied by the scanning gate microscopy (SGM).** Knowing the electric field distribution around the constricted region of the device at a given $V_b$ is important for interpreting the experimental results. For this sake we carry out the measurements of SGM:[37] We apply a voltage to the tip, $V_{tip}$, and measure the conductance of the device as a function of $V_{tip}$. In the fixed gate-bias condition, $V_{tip} = 0$, the current slightly changes, $\Delta I_{sd}$, depending on the location of the tip. This is because the local electron density below the tip changes, $\Delta n(r)$, due to the spatially varying local electrostatic potential $\phi(r)$ of the electron system. Extended data Fig.8 shows that $\Delta I_{sd}$ exhibits a sharp dip as the tip is scanned along the y-direction through the constricted region as shown by the red line for $V_b=9V$. The spatial profile of $\phi(r)$ can be obtained by mapping the gate bias voltage, $V_{tip,0}(r)$, at which the current is kept unchanged; viz., $\phi(r) = V_{tip,0}(r)$ for $\Delta I_{sd}= 0$. The data points of $V_{tip,0}$ in Extended data Fig.9 shows that $V_{tip,0}(y)$ or $\phi(y)$ rapidly increases by $\Delta V \approx 5.0$ V, from ~1.2V to ~6.2V, in a narrow range ($|y|<240$ nm) of the constriction. The electric field in the constricted region is hence evaluated to be $E_c \approx (5.0V)/(480nm) = 104$ kV/cm for $V_b = 9.0V$. The gradient of $V_{tip,0}$ (or the electric field E) outside this region is distinctly smaller: $E(y)$ at $y=500$nm is more than one order of magnitude smaller than $E_c$. It should be stressed that the hot electron distribution (Figs. 2e-g) expands significantly beyond the high-electric field region. The difference between $\Delta V \approx 5.0$ V and $V_b = 9.0$ V is ascribed to the voltage drop in the 2DEG region outside the constricted region including the 2DEG leads (see Methods section 'Estimation of the electric field').

**Estimation of the electric field.** The solid line in Extended data Fig.9 shows estimated values of the electric field, $E_c$, in the constricted region as a function of $V_b$. The estimation is made in two different manners. In the lower $V_b$-range (< 3.0V), where the transport is in the linear regime (Fig. 2h), $E_c$ is evaluated from $E_c = <v_d>/\mu$ with $\mu= 0.1668$ m$^2$/Vs, where $<v_d>= I/(n_{2D}eW_e)$ (see Methods section 'Important quantities relevant to the transport'). In the higher $V_b$-range (> 4.5V), where the current is saturated, $E_c$ is estimated through $E_c = \Delta V/L_c = (V_b - V_{lead})/L_c$, where $V_{lead}$ is the voltage drop in the 2DEG leads roughly approximated by $V_{lead} = 2\{(L_{lead}/W_{lead})/(n_{2D}e\mu)\}I$ with $L_{lead}= 190$ μm and $W_{lead}= 20$ μm and $L_c \approx 480$ nm is the effective length of the constricted 2DEG region (Extended data Fig.8). For $V_b = 9.0$ V, $\Delta V =V_b-V_{lead} =$ 5.76 V or $E_c \approx$ 120 kV/cm is derived, which is in substantial agreement with $\Delta V \approx$ 5.0 V or $E_c \approx$ 104 kV/cm obtained independently in the measurement of SGM (Extended data Fig.8). The uncertainty of the determination (by ca.15%), which can be partly ascribed to the presence of 2 DEG transition regions connecting the leads and the constricted 2DEG region, may not significantly affect the discussion in this work.

**Valley transfer.** In GaAs the bottom of the conduction band lies in the $\Gamma$ valley, and the second and the third local minima occur, respectively, in the upper satellite L- and X-valleys with the valley splitting energies $\Delta\varepsilon_{\Gamma L} \approx$ 320 meV and $\Delta\varepsilon_{\Gamma X} \approx$ 550 meV:[31] The density of states effective masses of electrons in the respective valleys are $m_L^D$= 0.56$m_0$ and $m_X^D$= 0.85$m_0$.[38] When the kinetic energy $\varepsilon$ of $\Gamma$-valley electrons exceeds $\Delta\varepsilon_{\Gamma L}$ or $\Delta\varepsilon_{\Gamma X}$, they transfer to the respective valleys at an intervalley scattering rate $1/\tau_{\Gamma \to L} \approx 1/(450 \text{ fs})$[39] or $1/\tau_{\Gamma \to X} \approx 1/(45 \text{ fs})$.[8]

The valley transfers lead to the Gunn effect in conductors of relatively long channels (>10 μm),[28] where the negative differential conductivity (NDC) gives rise to formation of a self-organized high-electric-field domain that propagates the conductor. In short channels (<1 μm), the Gunn effect is avoided because the self-organized high-electric-field domain is not formed. Instead, the valley transfer is suppressed due to a short transit time and the drift velocity of electrons exceeds the steady-state value of long channels ("velocity overshoot"). In our short-channel device, the population in the L-valleys can be neglected for simplicity because of the low intervalley scattering rate.[8,39] Existing experiment[8] and Monte Carlo simulation[29] make us to suggest that the relative population of X-valley electrons in the constricted region in this work starts to appreciably increase with increasing $E_c$ ($V_b$) above ca. 15kV/cm (3 V) and reaches a value larger than 75% in $E_c$ ($V_b$) above ca.60 kV (6 V) as shown in Extended data Fig.9.

Letting $\rho_{\text{n-GaAs},\Gamma}(\omega)$ and $T_{e,\Gamma}$ be the EM-LDOS and the effective electron temperature of the $\Gamma$-valley electrons, the excess noise intensity is roughly given by $V_{\text{SNoiM}} \propto \rho_{\text{n-GaAs},\Gamma}(\omega)[\hbar\omega/\{\exp(\hbar\omega/k_BT_{e,\Gamma})-1\}]$. As noted in Methods section 'Electromagnetic local density of states (EM-LDOS)', calculation shows $\rho_{\text{n-GaAs},e} \propto n^{1.5}$. This, together with the relation $\hbar\omega/\{\exp(\hbar\omega/k_BT_{e,\Gamma})-1\} \propto k_BT_{e,\Gamma} \propto \langle\varepsilon_\Gamma\rangle$ for the limit of high $T_{e,\Gamma}$, leads to $V_{\text{SNoiM}} \propto n_\Gamma^{1.5}\langle\varepsilon_\Gamma\rangle$ as a rough measure. In the text, we apply even more simplified assumption, $V_{\text{SNoiM}} \propto n_\Gamma^{1.0}\langle\varepsilon_\Gamma\rangle$, which does not substantially affect the discussion.

# Extended Data Figures

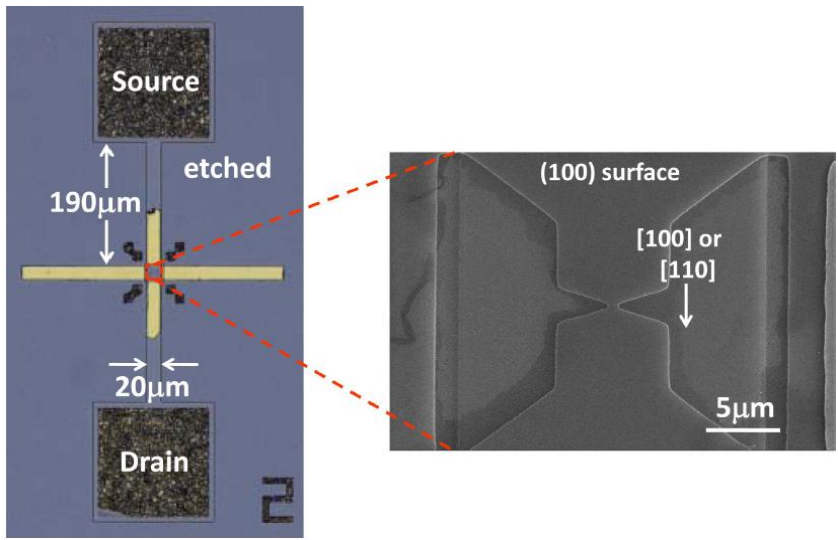

**Extended Data Fig.1 | The GaAs/AlGaAs quantum well structure.**

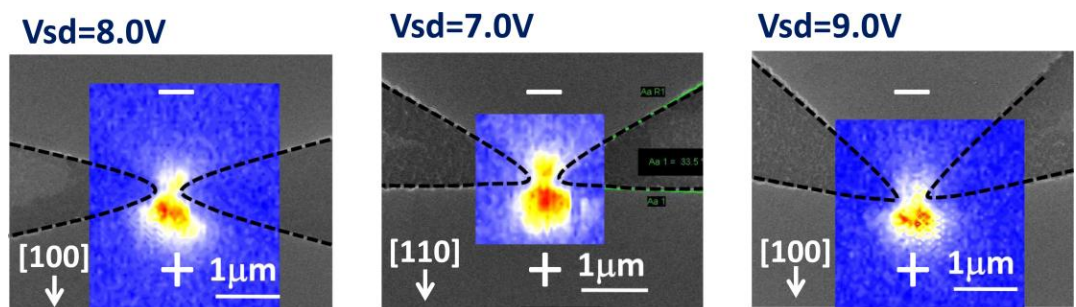

**Extended Data Fig.2| Nano-constriction device.**
Left; Optical microscope image of a GaAs/AlGaAs quantum well device. Right; Close-up of the central region with a constriction (SEM image):

**Extended data Fig.3 |Images of excess noise in different GaAs nano-constriction devices.**

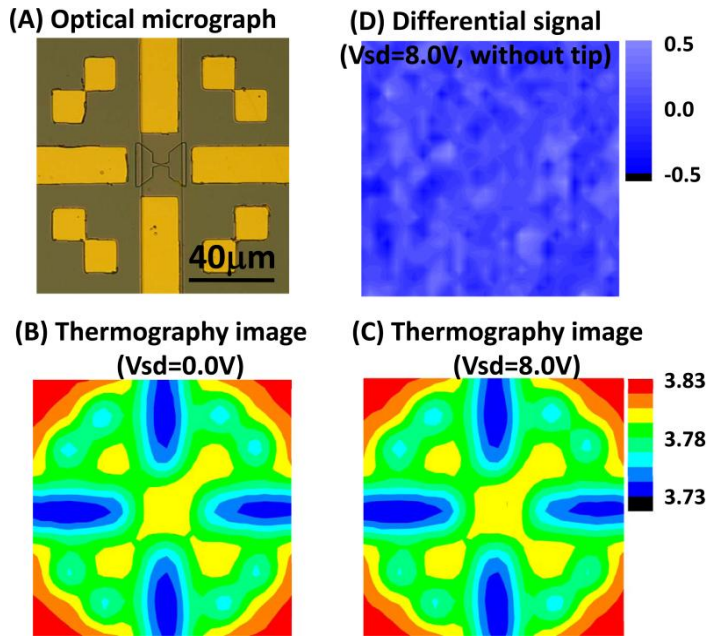

**Extended data Fig.4 | Thermography: Image of far-field radiation (14.1±0.5 μm).**
(A) Optical micrograph of a square of $(100~\mu m)^2$-area with a nano-constriction device at the center. (B)Thermography image of the same region as (A) without bias voltage, $V_b$= 0V. The contrast in the image arises not from temperature difference but from different emissivities of GaAs and Au. The spatial resolution is ≈ 15 μm. (C) Thermography image with the bias voltage of $V_b$= 8.0V. The image does not change appreciably from the image in the absence of bias voltage. (D) Image of difference between (B) and (C), taken by modulating $V_b$ at 5 Hz. The color scale is the same as those applied for Figs. 1c,d, Figs. 2a-f, Fig. 3a and Extended data Figs.3. In addition, the experimental condition in which the images of those figures are taken is similar to the one in which the image of (D) is taken, except that the tip is absent in (D). Hence overall (lattice) heating is not involved, and without the tip the radiation is not emitted from the nano-constriction device.

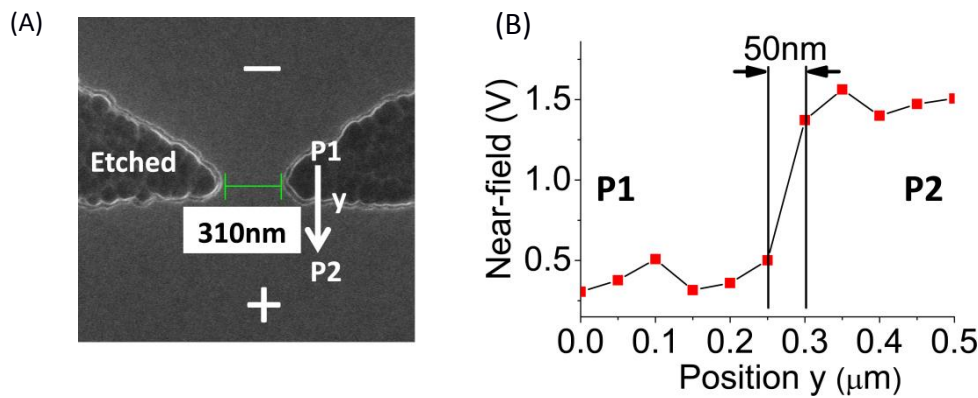

**Extended data Fig.5 | Spatial resolution of SNoiM.**
(A) Scanning Electron Microscope image of the GaAs nano-constriction device. (B)One-dimensional profile of the excess noise obtained by scanning the tip from position P1 to position P2 shown in (A) with $V_b$ = 7.0V. The spatial resolution, $\Delta X$, evaluated from the 10%-to-90% boundary transition is 50 nm.

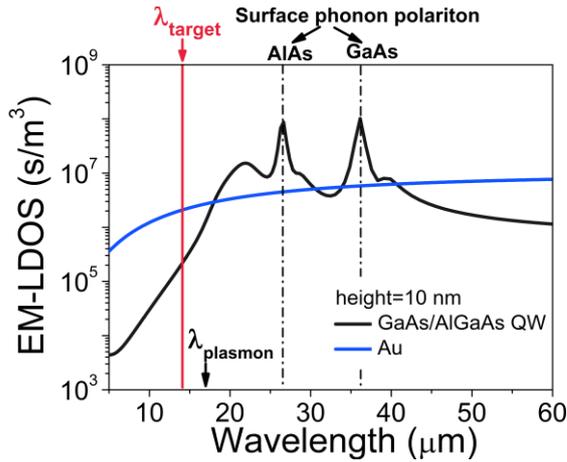

**Extended data Fig.6 | Electromagnetic local density of states (EM-LDOS).** Theoretically derived values for the GaAs/AlGaAs QW structure (Extended data Fig.1), $\rho_{n\text{-}GaAs}(h,\omega)$, and for gold, $\rho_{Au}(h,\omega)$, at height of h = 10 nm above the surface, are shown as a function of $\lambda = 2\pi c/\omega$. Two dominant peaks of $\rho_{n\text{-}GaAs}(h,\omega)$ are due to the surface phonon polariton resonances of GaAs and AlAs, respectively. The target wavelength, $\lambda_{target}$= 14.1 μm, and the plasma wavelength $\lambda_{plasma}= 2\pi c/\omega_p$ = 17.5 μm with $\omega_p = \{ne^2/(m_\Gamma^* \varepsilon_\infty)\}^{1/2} = 2\pi \times 17.1$ THz, n= 3.3 x $10^{24}$m$^{-3}$, $m_\Gamma^*$= 0.078 $m_0$ and $\varepsilon_\infty$= 10.9$\varepsilon_0$ ($\varepsilon_0$: the dielectric constant of vacuum) are indicated by the arrows (see Methods section 'Important quantities relevant to the transport' for parameters).

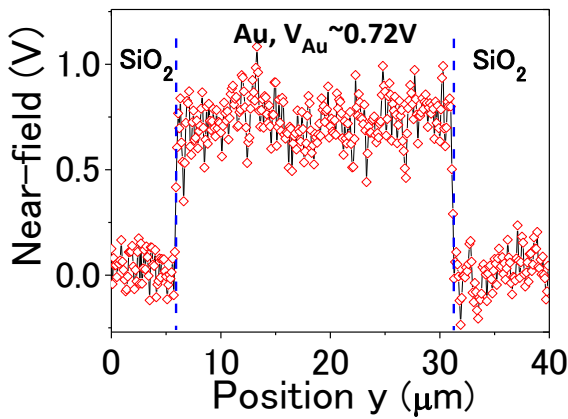

**Extended data Fig.7 | Thermal evanescent fields (thermal noise) on gold.**
A 25-μm wide gold stripe of 150nm-thickness is deposited on top of SiO$_2$ substrate. Without external stimulation, difference in the energy density of thermally activated electric and magnetic evanescent fields between SiO2 and Au is recorded in the one-dimensional profile at room temperature (300K). This data is compared with the excess noise data studied on the 2DEG system in GaAs nano-devices for the estimation of the effective electron temperature $T_e$. (Methods section 'Estimation of $T_e$'.)

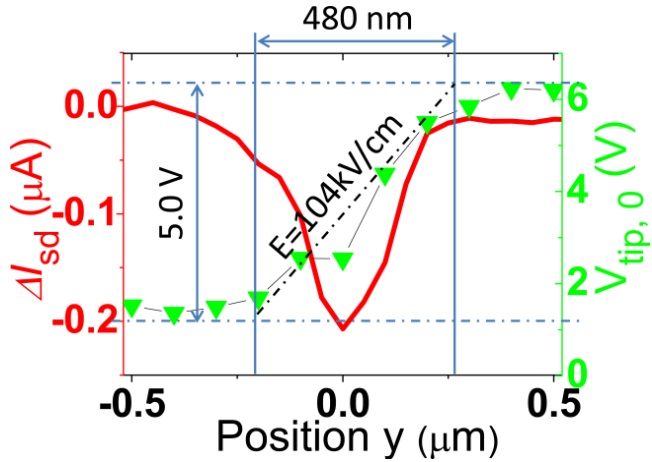

**Extended data Fig.8 |Electrostatic potential distribution in the constricted region.**
The red line shows the tip-induced change in current, $\Delta I_{sd}$, for $V_b$=9.0V as a function of y (Figs. 2) when the tip-bias voltage is fixed at zero ($V_{tip}$=0). The point of y=0 corresponds to the center of constricted region. Green triangles show $V_{tip,0}$; viz., the value of $V_{tip}$ at which $\Delta I_{sd}$= 0.

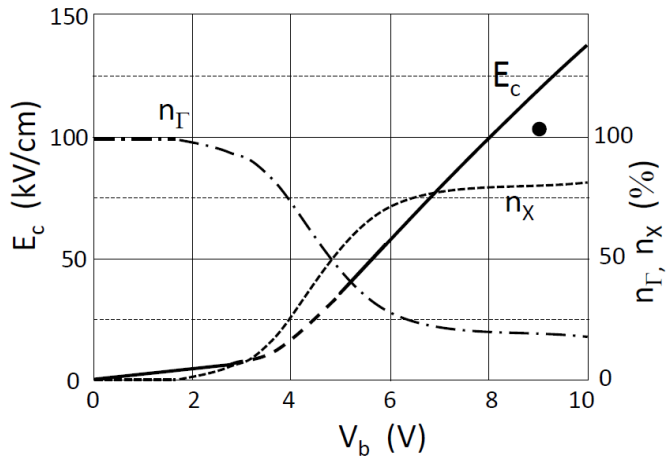

**Extended data Fig.9 |$E_c$, $n_\Gamma$ and $n_X$ against bias voltage $V_b$.** The solid lines indicate estimated values of the electric field $E_c$ at the constriction (see Methods section 'Estimation of the electric field'). The two solid lines for the lower- and the higher-$V_b$ ranges ($V_b$<3.0V and 4.5V< $V_b$) are smoothly connected by a broken line for guide for eyes. The black dot ($E_c$ =104 kV/cm at $V_b$ = 9.0V) shows a value determined by the SGM (Extended data Fig. 8). The dashed line and the dotted line show values of the relative population of electrons in the $\Gamma$- and the X-valleys in the constricted region, $n_\Gamma \equiv N_\Gamma/N$ and $n_X \equiv N_X/N$ with N the total density of electrons. The values are taken from theoretical values, $n_\Gamma^{cal}$, $n_L^{cal}$ and $n_X^{cal}$, derived as a function of electric field E by a Monte Carlo simulation considering long channels.[30] For our short-channel device, we ignore the contribution of L-valleys for simplicity and assume $n_\Gamma = n_\Gamma^{cal} + n_L^{cal}$ and $n_X = n_X^{cal}$. The values are plotted against $V_b$ by noting the relation $E_c(V_b)$.